\documentclass[conference]{IEEEtran}
\usepackage[utf8]{inputenc}
\usepackage{environ}
\usepackage{amsmath}
\usepackage{amssymb}
\usepackage{graphicx}
\usepackage{placeins}
\usepackage{adjustbox}
\usepackage[table,xcdraw,usenames,dvipsnames]{xcolor}
\usepackage{dsfont}

\usepackage{hyperref}

\definecolor{source}{HTML}{154360}
\definecolor{target}{HTML}{17A589}

\usepackage{makecell}

\NewEnviron{myequation*}{%
    \begin{equation*}
    \scalebox{1.2}{$\BODY$}
    \end{equation*}
    }

\NewEnviron{myequation}{%
\begin{equation}
\scalebox{1.2}{$\BODY$}
\end{equation}
}

\NewEnviron{myalign}{%
\begin{align*}
\scalebox{1.2}{\BODY}
\end{align*}
}

\ifCLASSINFOpdf
\else
\fi
\usepackage{algorithm, algpseudocode}%%%seems super usefull for non mac user !
\algrenewcommand\algorithmicrequire{\textbf{Input:}}
\algrenewcommand\algorithmicensure{\textbf{Output:}}

\usepackage{stfloats}
\begin{document}
\title{\vspace*{-0.5cm}  Pick the Largest Margin \\ for Robust Detection of Splicing}
%\title{On the Tradeoff between Robustness and Accuracy for Splicing Detection}

% Yo
\author{\IEEEauthorblockN{Julien SIMON de KERGUNIC}
\IEEEauthorblockA{Centrale Lille,\\ Cité Scientifique,\\ 59650 Villeneuve-d'Ascq, France\\
Email: julien.simon@centrale.centralelille.fr}

\and
\centering

% Yo
\IEEEauthorblockN{Rony Abecidan}
\IEEEauthorblockA{Univ. Lille, CNRS, Centrale Lille, \\ UMR 9189 CRIStAL,\\ F-59000 Lille, France\\
Email: rony.abecidan@univ-lille.fr}

%\IEEEauthorblockA{ Univ. Lille, CNRS, Centrale Lille, \\ Institut Mines-Télécom,\\ UMR 9189 CRIStAL,\\ F-59000 Lille, France\\
%Email: vincent.itier@imt-lille-douai.fr}

\and
\centering
\IEEEauthorblockN{Patrick Bas}
\IEEEauthorblockA{Univ. Lille, CNRS, Centrale Lille, \\ UMR 9189 CRIStAL,\\ F-59000 Lille, France\\
Email:  patrick.bas@cnrs.fr}

\and
\centering
\IEEEauthorblockN{Vincent Itier}
\IEEEauthorblockA{IMT Nord Europe, Institut Mines-Télécom,\\ Centre for Digital Systems, Univ. Lille, CNRS, Centrale Lille,\\ UMR 9189 CRIStAL, F-59000 Lille, France\\
Email: vincent.itier@imt-nord-europe.fr}

}

%Telephone: (800) 555--1212, Fax: (888) 555--1212}



\makeatletter
\newcommand{\linebreakand}{%
  \end{@IEEEauthorhalign}
  \hfill\mbox{}\par
  \mbox{}\hfill\begin{@IEEEauthorhalign}
}
\makeatother

\maketitle

%INCLUDES COPYRIGHT NOTICE: one of three copyright notice should be included.
%Uncomment the appropriate line below, according to the authors %affiliation:
\begin{figure}[b]
  % \vspace{-0.3cm}
  \parbox{\hsize}{\em
  %information about the event:
  WIFS`2024, December, 2-5, 2024, Roma, Italy.
  %copyright notice: one of four copyright notices below should be included. Choose the right one below according to the authors affiliation:
  %XXX-X-XXXX-XXXX-X/XX/\$XX.00 \ \copyright 2017 European Union.
  %XXX-X-XXXX-XXXX-X/XX/\$XX.00  \ \copyright 2017 Crown.
  %U.S. Government work not protected by U.S. copyright.
  %XXX-X-XXXX-XXXX-X/XX/\$XX.00 \ \copyright 2017 IEEE.
  XXX-X-XXXX-XXXX-X/XX/\$XX.00 \ \copyright 2024 IEEE.
  }\end{figure}

\newcommand\NPB[1]{{\color{green!70!black}{ \small { [PB: ~#1~]}}}}

\newcommand\NRA[1]{{\color{red!70!black}{ \small { [RA: ~#1~]}}}}

\newcommand\NVI[1]{{\color{blue!70!black}{ \small { [VI: ~#1~]}}}}

\newcommand\NJB[1]{{\color{magenta!70!black}{ \small { [JB: ~#1~]}}}}

\newcommand\QF[1]{\textsc{qf}#1}

\vspace*{-1.3cm}
\begin{abstract}
  Despite advancements in splicing detection, practitioners still struggle to fully leverage forensic tools from the literature due to a critical issue: deep learning-based detectors are extremely sensitive to their trained instances. Simple post-processing applied to evaluation images can easily decrease their performances, leading to a lack of confidence in splicing detectors for operational contexts. In this study, we show that a deep splicing detector behaves differently against unknown post-processes for different learned weights, even if it achieves similar performances on a test set from the same distribution as its training one. We connect this observation to the fact that different learnings create different latent spaces separating training samples differently. Our experiments reveal a strong correlation between the distributions of latent margins and the ability of the detector to generalize to post-processed images. We thus provide to the practitioner a way to build deep detectors that are more robust than others against post-processing operations, suggesting to train their architecture under different conditions and picking the one maximizing the latent space margin. %Additional resources will be available upon acceptance.
\end{abstract}

% Additional resources are available at \textcolor[HTML]{156199}{\href{https://github.com/RonyAbecidan/LeveragingGeometrytoMitigateCSM}{this link}}


% \fi
\IEEEpeerreviewmaketitle
\vspace*{-0.1cm}

\section{Introduction}
\label{sec:intro}

Splicing is the process of altering an image by taking objects from a different image and inserting them into the original, effectively changing its meaning or message.
Modern splicing detectors leverage deep architectures harnessing noise anomalies to perform their detections~\cite{bayar, catnet, trufor}, however their effectiveness in real-world scenarios often do not match the performances reported in the literature. One common explanation of this gap of performance is attributed to unknown post-processing transformations applied to the spliced images~\cite{fcs1,fcs2}. A simple post-processing pipeline applied on all images under scrutiny is indeed enough to create a significant domain shift, altering the distribution of both pristine and manipulated images and disturbing detectors trained by forensic analysts. This generalization problem is due to the fact that post-processing operations (e.g. sharpening, denoising, resizing etc.) alter the noise distribution of images while creating dependencies between pixels of both original and forged areas.

\begin{table}[ht]
  \vspace*{-0.9cm}
  \centering
  \begin{adjustbox}{max width=\linewidth}
    \begin{tabular}{|c|c|c|c|}
    \hline
    Seed & Source accuracy & Mean target accuracy &  Std. of Target Accuracy \\ \hline
    4 & 84\% & 72\% & 1.8 \\ \hline
    6 & 84\% & 74\% & 2.0 \\ \hline
    8 & 84\% & 66\% & 2.3 \\ \hline
    \end{tabular}
  \end{adjustbox}
  \caption{\scriptsize Impact of different initialization seeds on the out-of-domain performance of the Bayar detector \cite{bayar}. The mean target accuracy is computed by averaging the test accuracies of a Bayar Detector trained with three distinct seeds on 20 post-processed targets processed with RawTherapee. (Train) $N_{source} \sim 20,000$ patches;\\ (Test) $N_{source} \sim N_{target} \sim 7,000$ patches.
  }.
  \label{tab:teasing}
  \vspace*{-1cm}
\end{table}
Postprocessing consequently makes forgeries less noticeable for forensic detectors designed to detect statistical anomalies between pixels. Hence, all forgery detectors may be affected by post-processing.

\subsection{Robust Detection against Post-Processing: Prior Arts}
In forensic literature, the \textit{domain shift} caused by post-processing pipelines is present in various manipulation detection problems, including photo-editing \& watermarking detection, and steganalysis~\cite{fcs1, watermark, giboulotcsm}. Several solutions, often inspired by the machine learning literature, have been proposed to address this issue.

- \textit{Data-centric} approaches focus on searching or building relevant training sets allowing any detector to generalize across multiple distributions which are different from the training domain. The simplest method involves artificially augmenting the training set by randomly applying standard post-processing operations (denoising, sharpening, JPEG compression, etc.) to the source images~\cite{domainrandom2}. However, recent studies in steganalysis suggest that it is more effective to carefully select the operations to apply rather than using a random mixture of operations~\cite{clevermixture,wifs2022}. When target images are available, other strategies propose to select relevant sources for the target or to estimate the post-processing pipeline applied to the target to reproduce training samples following the target distribution~\cite{doublecompression,wifs2023}. This allows to train a detector on a source very related to the target. A common issue with data-centric approaches is the uncertainty that the constructed or selected datasets adequately cover all target domains of interest.

- \textit{Detector-centric} approaches aim to build or update detectors to be more robust against out-of-distribution data. This is particularly relevant in domain adaptation, where a model trained on a labeled source needs to generalize to an unlabeled target. The most famous domain adaptation strategy for image forgery detection is ForensicTransfer~\cite{forensictransfer}. This architecture leverage an autoencoder to learn a latent space separating domain-specific information irrelevant for detection, from domain-invariant information relevant for detection. This approach has shown promising results for robust synthetic image detection when a few labels from the target set are available. Classical strategies from machine learning literature are also used to find feature-invariant spaces with a relevant adaptation cost~\cite{ganin,long}, typically a distance between distributions or an adversarial loss added to the binary cross-entropy loss~\cite{advsforensics,wifs2021}. While detector-centric strategies are promising, some require labeled target data and others assume a balanced distribution of pristine and manipulated images in their targets. Moreover, models adapted to specific targets tend to overfit and do not generalize well to other targets, requiring practitioners to retrain multiple models for different investigations.

To our knowledge, there is a clear lack of studies on the out-of-distribution robustness of manipulation detectors in scenarios where no target data is available at training time. We highlight this issue by presenting an interesting phenomenon in Table~\ref{tab:teasing}, which is never mentioned in forensic literature. This table displays the performance of the same architecture trained starting from three different seeds, showing similar performance on the testing set of the source, while exhibiting very different performances on 20 different post-processed targets. We highlight here that different trainings of the same architecture do not lead to the same robustness against post-processing, despite similar performance on the source (convergence towards different local minimums). This disparity raises important questions about the best practices for training to ensure consistent robustness against different  post-processing attacks.

\vspace*{-0.1cm}
\subsection{Contributions}

Our primary goal is to understand why the same architecture dedicated to splicing detection can exhibit different behaviors when faced with post-processing attacks. We consider a realistic scenario where a practitioner can train several detectors in various ways and needs to identify the most robust detector for investigations, without targeting a specific post-processing. Our study aims to:

\begin{itemize}
  \item Explore the factors that make a splicing detector robust to post-processing attacks.
  \item Derive best practices for forensic practitioners to enhance the performance of their detectors on scrutinized images.
\end{itemize}

This paper is the first to address the challenge of post-processing domain shift in forensics by proposing multiple trainings of the same architecture. Our contributions are listed as follow:

\begin{enumerate}
  \item We highlight that highly specific training on a source results in poor generalization performance on post-processed targets.
  \item We demonstrate a clear correlation between the separation of training data into pristine and spliced classes within first and last latent spaces and the ability of a detector to generalize to post-processed images.
  \item We compare the impact of pooling and normalization layers on the ability of the detector to generalize to a variety of post-processed samples.
\end{enumerate}
 
The structure of the paper is as follows:
Section~\ref{sec:formalization} presents the formalization of our objective and introduces the notion of latent space margins.
In Section~\ref{sec:margin}, a series of experiments are conducted to gain a deeper understanding of factors contributing to a robust learning for a splicing detector.
The influence of pooling and normalization operators on the robustness of detection is notably examined in this section. Lastly, Section~\ref{sec:conclusion} serves as the conclusion, summarizing the main findings and contributions of this research and proposing some perspectives.

%\newpage

\section{Formalization}
\label{sec:formalization}
\subsection{Problem formulation and scenario}
In accordance with \cite{sepak}, we define a processing pipeline as a vector $\omega \in \Omega$ that encompasses all the parameters associated with the pipeline, such as the downsampling factor, the denoising coefficient, the JPEG quality factor, \textit{etc}. For splicing detection, machine learning models are commonly used:
\vspace*{-0.2cm}
\begin{align*}
    f(x \mid \theta_{\omega}) : \ & \mathcal{X} \rightarrow \{pristine,manipulated\} \\
    & x \mapsto y
  \end{align*}

Here, $\theta_{\omega} \in \Theta$ represents the learned parameters using pristine and spliced images post-processed using parameters $\omega$. To assess the impact of post-processing mismatch, it is common to compute the \textit{generalisation gap} between a source $s$ (training base) and a target $t$ (evaluation base):
\vspace*{-0.5cm}
  {\normalsize
  
  \begin{align*}
    \mathcal G_{f(x \mid \theta_{\omega})}(\omega_s,\omega_t) = 
    \ &\mathbb E_{(x,y) \sim P((x,y)| \omega_s)} (f(x \mid \theta_{\omega_s} ) = y) \\ - \
     &\mathbb E_{(x,y) \sim P((x,y)| \omega_t)} (f(x \mid \theta_{\omega_s} ) = y) \ \ (1). 
  \end{align*} 

  }
This gap represents the difference of performance between the ideal scenario where the post-processing of the target is the same as the one of the source and the real scenario where we do not know the post-processing of target images.

Here we do not have access to target samples at training time. Ideally, we want to build a detector as robust as possible against unknown post-processings. We assume that a forensic practitioner is familiar with a splicing detector but does not know what to do to make it robust.

\subsection{Latent spaces margins}

In splicing detection, we have two classes: \textit{pristine} and \textit{spliced}. Accordingly, our models output two logit scores, \( f_1 \) and \( f_2 \), for each input \(x \in \mathcal{X}\). The class with the highest score is chosen as the predicted label, given by \( i^* = \arg \max_i f_i(x) \). Deep detectors are made of successive layers, with each layer projecting its input into a new latent space\footnote{A latent space is a lower-dimensional representation of data capturing its underlying structure and features.}. The linear decision boundary in the final latent space appears non-linear in previous latent spaces, leading to a distinct decision boundary for each latent space. The decision boundary $\mathcal{D}^l$ of the \( l \)-th latent space of our detector is the set of points in this latent space $x^l$ where the detector is uncertain between the two classes:
$$
\mathcal{D}^l = \left\{ x^l \mid f_1(x^l) = f_2(x^l) \right\}. \ \ (2)
$$
We can now define the margin of a latent sample \(x^l\) with respect to this latent boundary $\mathcal{D}^l$ as the smallest perturbation \(\delta^l\) necessary to move \(x^l\) to the decision boundary of the \( l \)-th latent space:
\[
d^p_{f, x^l} = \min_{\boldsymbol{\delta}} \|\delta^l\|_p \quad \text{s.t.} \quad f_1(x^l + \delta^l) = f_2(x^l + \delta^l) \ \ (3).
\]

The performance gap caused by post-processing mismatches exists because deep splicing detectors tend to learn biases that are very specific to the training distribution.
This creates decision boundaries suited to source samples but ineffective for target samples with different distributions. Ideally, we feel that a general learning would space out samples from different classes more clearly in the latent spaces. Indeed, a boundary too close to source samples means that even slight variations or noise addition in the training data (such as those introduced by post-processing pipelines) can cause the new data from another distribution to overshoot this boundary, leading to misclassifications. A previous research showed a correlation between the generalization gap and the distribution of latent margins (distances between latent decision boundaries and the training points from each class) \cite{margin}. However, this study tested this intuition with only two target distributions and used image classification classifiers relying on semantics for their decisions. In this paper, we propose to check the validity of this correlation in the context of splicing detection. We validate this rational using with twenty targets that have undergone various post-processings, using the Bayar detector \cite{bayar}, which bases its decisions on image noise rather than semantics.

% ForensicTransfer créé justement des grosses margin, contrastive learning aussi.

% Maximizing the margin (the distance between the decision boundary and the closest data points from each class) ensures that even small variations or noise in the input data do not easily cause misclassifications.

% En contrastive learning, l'idée est précisément de construire des espaces où les classes sont très bien séparés. 

\section{Best training practices for Robust Detection}
\label{sec:margin}
\subsection{Experimental protocol}

\subsubsection{Detector's choice and hyperparameters}

\noindent
For our experiments, we use the popular forgery detector developed by Bayar and Stamm~\cite{bayar}. This deep detector is known for its simple yet effective design, making it an interesting choice for standard databases and sufficient for our analysis. Its architecture is traditional (Convolution + Max Pooling + Fully Connected Layers). However, the very first convolutional layer is constrained to perform high-pass filtering: 

\begin{center}
	\vspace{-0.4cm}
	\footnotesize{
		$$
\left\{\begin{array}{l}
\boldsymbol{w}_{k}^{(1)}(0,0)=-1, \\
\sum_{m, n \neq 0} \boldsymbol{w}_{k}^{(1)}(m, n)=1.
\end{array}\right.
$$
	}
\end{center}
\noindent
This constraint fosters the extraction of relevant low-level forensic features The other layers are non-constrained and act as usual.  The following choices were made concerning the optimization 
strategy and the hyperparameters:
\begin{itemize}
	\item The maximal number of epochs is fixed at 115, a reasonable amount of epochs enabling to observe a convergence in practice.
	\item The optimizer is SGD.
	\item The batch size is fixed at 128, a reasonable size for computation on a regular GPU while ensuring a good convergence of our detector.
	\item The learning rate (lr) is fixed to $10^{-3}$ with a lr scheduler dividing by 10 the lr with a patience of 4 epochs.
	\item The initialization of our weights is the one by default
	on pytorch~\cite{he2015delving}. For each study, we initialized our forgery detector with the common seed 22 in order to make our results reproductible and ensuring
	a fair comparison of them.
\end{itemize}

We cut source and targets into train/validation/test with
the proportion $0.6/0.2/0.2$. To get the best of our model and avoid overfitting, we consider for each experiment an early stopping callback based on the accuracies obtained on the source validation set.

\begin{figure}[h]
  \centerline{
  \includegraphics[width=\columnwidth]{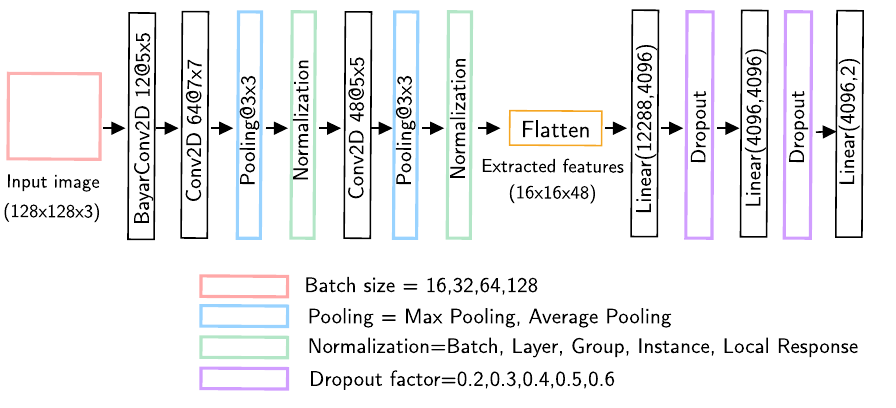}}
    \vspace*{-0.2cm}
  \caption{Scheme of Bayar Detectors \cite{bayar}. Colored cells indicates the hyperparameters or operators that are changing over our 200 trainings.}
  \label{fig:bayar}
\end{figure}

\subsubsection{Construction of source and target domains}

The largest public dataset dedicated to splicing detection is DEFACTO~\cite{DEFACTODataset}. We have chosen this dataset for our analysis due to several key factors: its extensive size, the high-quality and realistic images it contains, and the full control it offers over the post-processing pipelines, with images saved in TIF format. These features make DEFACTO an ideal choice for our study. 

To prepare source and targets datasets for our experiments, we start by dividing the splicing category of DEFACTO into two equally-sized, independent sets: one for the source and one for targets. Each image is then cut into $128 \times 128$ patches to ensure a uniform training process. For each patch from both source and target bases, we create a \textit{spliced} class selecting patches with a tampered surface ratio between 10\% and 40\% of their total area. This specific range is carefully chosen to balance two factors:  if the tampered area is too small or too high, the detector may struggle to differentiate two noise distributions. By selecting carefully our patches, we aim to create a realistic and challenging environment for evaluating the performance of our splicing detectors. Based on the number of spliced patches, \textit{pristine} patches are then selected in equal quantity to constitutes
balanced classes. This preprocessing gives us around 20.000 patches for our training sets and 7.000 patches for our testing sets. To build our source, we only work with original TIF images of DEFACTO. For the targets, post-processing are applied. All these processing are done on the original TIF images before cutting them into patches to prevent artifacts. Note that patches from two distinct post-processed targets are similar in terms of content. This is done to uniquely attribute the observed generalization gap to the application of the post-processing pipeline and not to the content associated to the training or testing sets.
\subsubsection{Post-Processing Pipelines} 

We created a set of 20 processing pipelines playing with  Wavelet Denoising and Sharpening operations of \href{https://www.rawtherapee.com/}{RawTherapee}, an open source software for image processing. We apply JPEG compression with a quality factor of 70 at the end of each pipeline using  \href{https://www.imagemagick.com/}{Imagemagick} to fully control it. Our choice of pipelines was a good tradeoff between target realism and the observation of strong generalization gaps. Details about these post-processing pipelines are presented in Figure \ref{fig:pipelines}.

\begin{figure}[h!]
  \centerline{
  \includegraphics[width=\columnwidth]{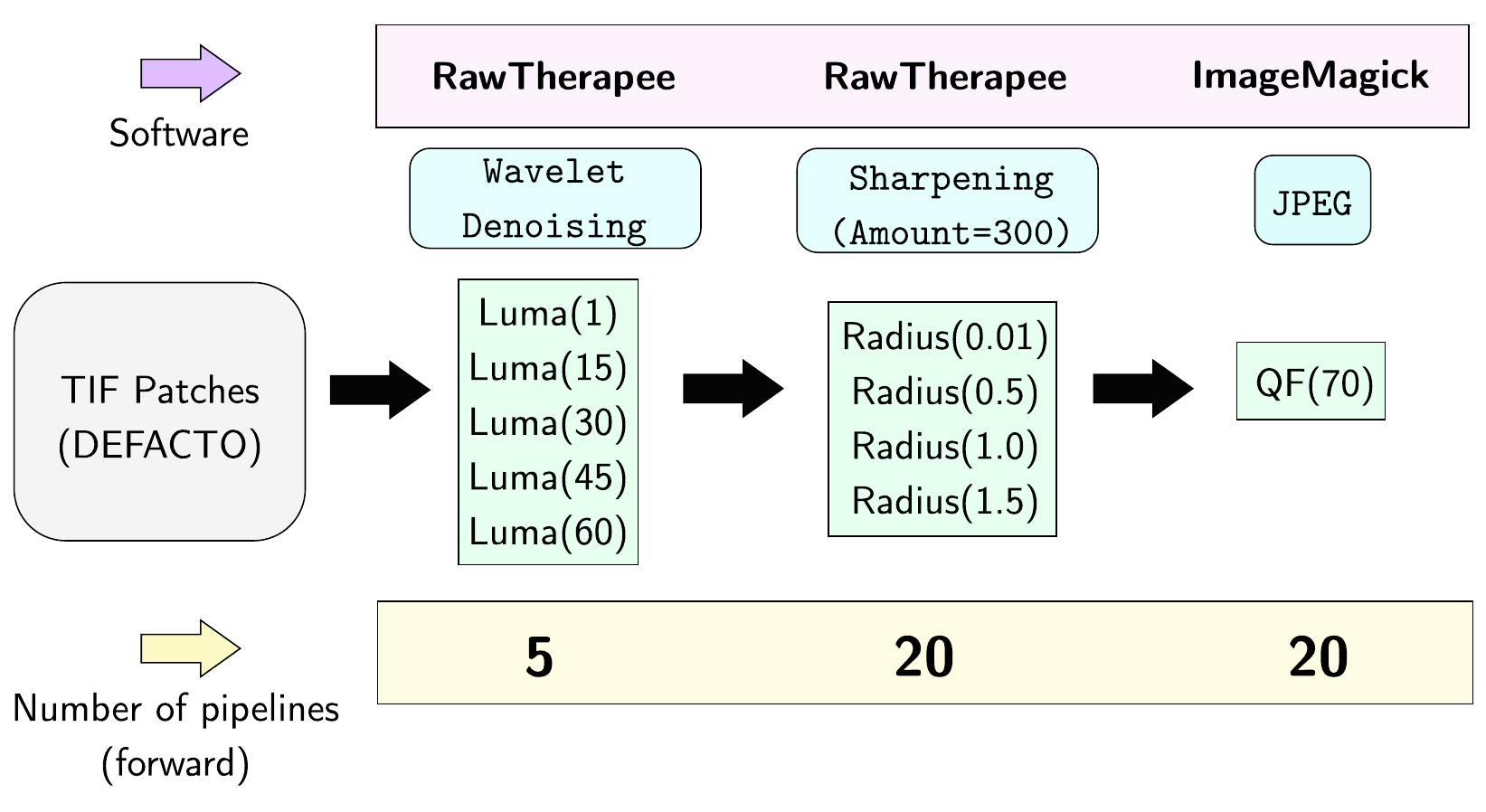}
    }
    \vspace*{-0.2cm}
  \caption{Details about the  post-processing pipelines applied to the target TIF set for our study.}
  \label{fig:pipelines}
\end{figure}

\subsubsection{Multiple trainings of the same architecture}
We propose here to play with hyperparameters and operators  of the Bayar detector to study how they influence its ability to generalize to post-processed samples. We perform 200 trainings of the Bayar Detector modifying batch size, pooling, normalization and drop out following the scheme of Figure \ref{fig:bayar}.
\subsubsection{Quantile plots for analysis} Quantile plots help us visualize how the generalization gap distribution
evolves within a sliding window centered in successive values of diverse metrics. For each quantile plot, we scan metric points with a given step and a given window size (precised in captions), balancing the tradeoff between localization and accuracy on quantiles computation. 

\subsection{Source overfitting through the epochs}

The results obtained from our 200 trainings validate the issue of source overfitting briefly mentioned in \cite{margin}: as the accuracy on the source test set increases, the generalization gap across all target domains also increases. We believe this happens because the network learn more and more source-specific features over the epochs, ultimately focusing on specific biases present in the source samples to enhance its performance on this source. This observation is illustrated in Figure \ref{fig:source_overfitting}. It shows how the generalization gap evolves w.r.t. the final accuracy on the source. Therefore, while forensic analysts should strive for good performance on their source, we do not recommend excessive trainings of splicing detectors if they have to evaluate their detectors on unknown targets.  
\begin{figure}[h]
  \centerline{
  \includegraphics[width=0.8\columnwidth]{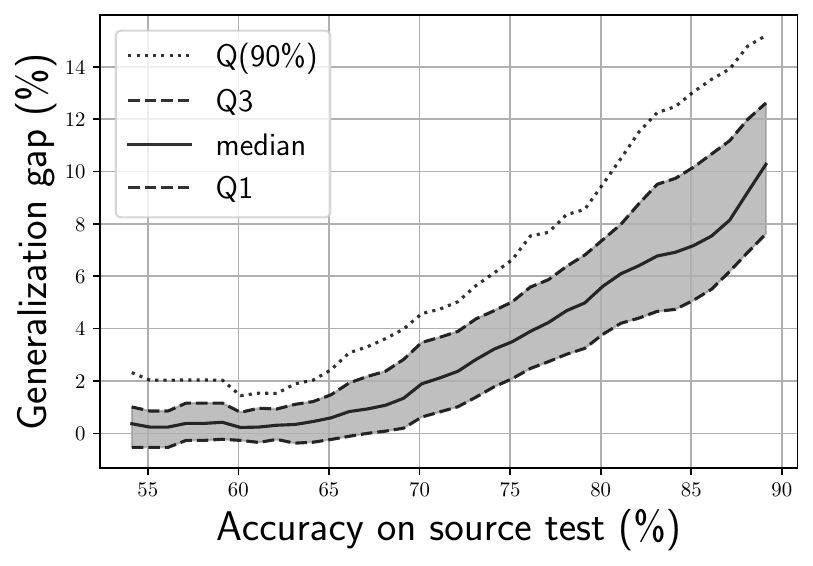}
    }
    \vspace*{-0.2cm}
  \caption{Quantile plots representing the evolution of the generalization gap over our 20 domains according to the accuracy of our detectors on the source test. Q1 is the first quartile, Q3 is the third quartile and Q(90\%) is the 90th percentile. Metric points are scanned with a step of 1 and a given window size of 10.}
  \label{fig:source_overfitting}
\end{figure}
\subsection{Latent margins and generalization gaps}

\begin{figure}[ht!]
  \centerline{
  \includegraphics[width=\columnwidth]{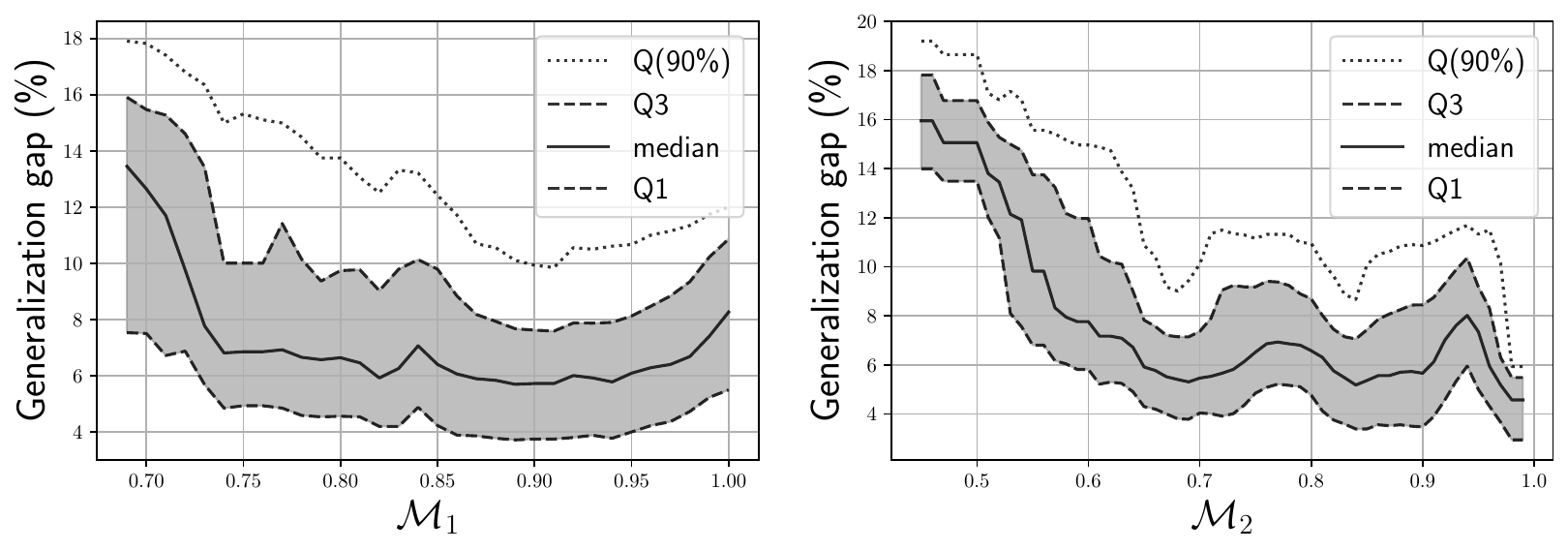}
    }
    \vspace*{-0.3cm}
  \caption{Quantile plots representing the evolution of the generalization gap over our 20 domains according to the margin metrics $\mathcal M_1$ and $\mathcal M_2$ computed using latent margins from all layers. These metrics are normalized for comparison. Q1 is the first quartile, Q3 is the third quartile and Q(90\%) is the 90th percentile.Metric points are scanned with a step of 0.01 and a given window size of 0.1. \\ \hspace*{2.5cm}}
  \label{fig:margin_gp}
\end{figure}

\vspace*{-0.4cm}
Here we want to check if there is a correlation between distances to the boundary of well-classified latent samples and generalization gaps. Given that all our models did not converge equally, we restrict our studies to the 138 models achieving an accuracy of at least 75\% on the source, in order to discard low generalization gaps caused by underfitting. For the computation of latent margins, we follow a methodology inspired by \cite{margin}:
\begin{enumerate}
  \item Estimate the latent margins $d^2_{f, x^l}$  of each source sample $x^l$ using logits $f_1,f_2$ and their gradients w.r.t. each layer, while taking care of normalizing them for scale independence. More details about this computation are available in \cite{margin} and our github repo. Following \cite{margin}, we discard negative margins caused by misclassifications.
  \item Summarize margins distributions per latent space with vectors $\boldsymbol{\mu_l}$ of descriptive statistics (first and third quartiles, median, upper and lower fences). Eventually, one could mix all the $\boldsymbol{\mu_l}$ in one vector $\boldsymbol{\mu}$.
  \clearpage
  \item Combine the margin statistics of every vector to derive a margin metric $\mathcal M$. We suggest testing straightforward combinations by computing the sum of the statistics raised to a certain power $\alpha$ : $\mathcal M_{\alpha} =  \sum_i \mu_i^{\alpha}$. Raising $\alpha$ enable to better emphasize margin differences among the architecture.
\end{enumerate}

Contrary to \cite{margin}, our goal is not to predict the generalization gap with margin statistics but rather provide practitioners a metric enabling to assess the quality of their training for robust splicing detection.
\noindent
For each Bayar Detector $f^i(x|\theta_{s})$ trained on our source $s$ and each target $t$ post-processed with pipeline $\omega_t$, we produce couples $[\mathcal{M}_{\alpha}[f^i(x|\theta_{s})] \ , \ \mathcal G_{f^i(x|\theta_{\omega_{s}})}(\omega_{s},\omega_t)]$. Such couples disclose the eventual presence of a correlation between the $\mathcal{M}_{\alpha}$ and  the generalization gaps over 2760 points. Quantile plots of Figure \ref{fig:margin_gp} confirm that most robust detectors against post-processing attacks are the ones separating the most their latent training samples (especially with the use of $\mathcal{M}_{2}$). 

\subsection{Importance of the margin in each latent space}
\begin{table*}[b!]
  \centering
  \begin{adjustbox}{max width=0.6\linewidth}
  \begin{tabular}{|l|c|c|c|c|c|c|c|}
  \hline
   \cellcolor[HTML]{ffffe0} $\boldsymbol{\mathsf{Operator \ name}}$ & \cellcolor[HTML]{ffffe0} $\boldsymbol{\mathsf{Median \ Source \ Accuracy}}$ & \cellcolor[HTML]{ffffe0} $\boldsymbol{\mathsf{Median \ \mathcal M_2}}$& \cellcolor[HTML]{ffffe0}$\boldsymbol{\mathsf{Min \ \mathcal G}}$ & \cellcolor[HTML]{ffffe0}$\boldsymbol{\mathsf{Q1 \ \mathcal G}}$ &\cellcolor[HTML]{ffffe0}$\boldsymbol{\mathsf{Median \ \mathcal G}}$&\cellcolor[HTML]{ffffe0} $\boldsymbol{\mathsf{Q3 \ \mathcal G}}$ &\cellcolor[HTML]{ffffe0}$\boldsymbol{\mathsf{Max \ \mathcal G}}$\\ \hline
   \multicolumn{8}{|c|}{\cellcolor[HTML]{d3d3d3} {$\mathsf{Normalization}$}} \\ \hline
  $\mathtt{instancenorm}$ & \cellcolor[HTML]{8FF1B1} 80\% & \cellcolor[HTML]{8FF1B1} 0.83 & \cellcolor[HTML]{8FF1B1} 0\% & \cellcolor[HTML]{8FF1B1} 2\% & \cellcolor[HTML]{8FF1B1} 3\% & \cellcolor[HTML]{8FF1B1} 4\% & \cellcolor[HTML]{8FF1B1} 6\% \\ \hline
  $\mathtt{batchnorm}$ & 82\% & 0.66 & 1\% & 7\% & 10\% & 13\% & 21\% \\ \hline
  $\mathtt{layernorm}$ & 82\% & 0.82 & \cellcolor[HTML]{8FF1B1} 0\% & 4\% & 6\% & 7\% & 11\% \\ \hline
  $\mathtt{local\_response\_norm}$ & 83\% & 0.63 & \cellcolor[HTML]{8FF1B1} 0\% & 4\% & 6\% & 7\% & 16\% \\ \hline
  $\mathtt{group}$ & 85\% & 0.81 & \cellcolor[HTML]{8FF1B1} 0\% & 5\% & 7\% & 9\% & 15\% \\ \hline
  \multicolumn{8}{|c|}{\cellcolor[HTML]{d3d3d3} \textbf{$\mathsf{Pooling}$}} \\ \hline
  $\mathtt{average\_pooling}$ & \cellcolor[HTML]{8FF1B1} 82\% & \cellcolor[HTML]{8FF1B1} 0.83 & \cellcolor[HTML]{8FF1B1} 0\% & \cellcolor[HTML]{8FF1B1} 4\% & 7\% & 10\% & \cellcolor[HTML]{8FF1B1} 15\% \\ \hline
  $\mathtt{max\_pooling}$ & 83\% & 0.69 & \cellcolor[HTML]{8FF1B1} 0\% & 5\% & \cellcolor[HTML]{8FF1B1} 6\% & \cellcolor[HTML]{8FF1B1} 8\% & 21\% \\ \hline
  \multicolumn{8}{|c|}{\cellcolor[HTML]{d3d3d3}  \textbf{$\mathsf{Dropout}$}} \\ \hline
  $\mathtt{dropout(0.2)}$ & \cellcolor[HTML]{8FF1B1} 82\% & 0.77 & \cellcolor[HTML]{8FF1B1} 0\% & \cellcolor[HTML]{8FF1B1} 4\% & \cellcolor[HTML]{8FF1B1} 6\% & \cellcolor[HTML]{8FF1B1} 7\% & \cellcolor[HTML]{8FF1B1} 15\% \\ \hline
  $\mathtt{dropout(0.3)}$ & 83\% & 0.76 & \cellcolor[HTML]{8FF1B1} 0\% & \cellcolor[HTML]{8FF1B1} 4\% & \cellcolor[HTML]{8FF1B1} 6\% & 9\% & 16\% \\ \hline
  $\mathtt{dropout(0.4)}$ & \cellcolor[HTML]{8FF1B1} 82\% & 0.77 & \cellcolor[HTML]{8FF1B1} 0\% & \cellcolor[HTML]{8FF1B1} 4\% & \cellcolor[HTML]{8FF1B1} 6\% & 8\% & 16\% \\ \hline
  $\mathtt{dropout(0.5)}$ & 83\% & 0.76 & \cellcolor[HTML]{8FF1B1} 0\% & \cellcolor[HTML]{8FF1B1} 4\% & 7\% & 10\% & 20\% \\ \hline
  $\mathtt{dropout(0.6)}$ & 83\% & \cellcolor[HTML]{8FF1B1} 0.79 & \cellcolor[HTML]{8FF1B1} 0\% & 5\% & 8\% & 11\% & 21\% \\ \hline
  \multicolumn{8}{|c|}{\cellcolor[HTML]{d3d3d3} \textbf{$\mathsf{Batch Size}$}} \\ \hline
  $\mathtt{16}$ & 84\% & \cellcolor[HTML]{8FF1B1} 0.79 & 1\% & 5\% & 7\% & 10\% & 20\% \\ \hline
  $\mathtt{32}$ & 83\% & 0.78 & 1\% & 5\% & 7\% & 10\% & 18\% \\ \hline
  $\mathtt{64}$ & 80\% & 0.76 & \cellcolor[HTML]{8FF1B1} 0\% & 4\% & 6\% & 8\% & 21\% \\ \hline
  $\mathtt{128}$ & \cellcolor[HTML]{8FF1B1} 77\% &  0.69 & \cellcolor[HTML]{8FF1B1} 0\% & \cellcolor[HTML]{8FF1B1} 3\% & \cellcolor[HTML]{8FF1B1} 5\% & \cellcolor[HTML]{8FF1B1} 6\% & \cellcolor[HTML]{8FF1B1} 11\% \\ \hline
  \end{tabular}
  \end{adjustbox}
  \vspace*{0.2cm}
  \caption{Contribution of Normalization, Pooling, Dropout, and Batch Size. $\mathcal M_2$ is normalized similarly as in Figure \ref{fig:margin_gp}. The lowest source accuracy, the highest margins and the lowest generalization gaps are highlighted in green.}
  \label{tab:merged}
  \vspace*{-2.18cm}
\end{table*}
\begin{figure*}[t!]
  \centerline{
  \includegraphics[width=0.65\textwidth]{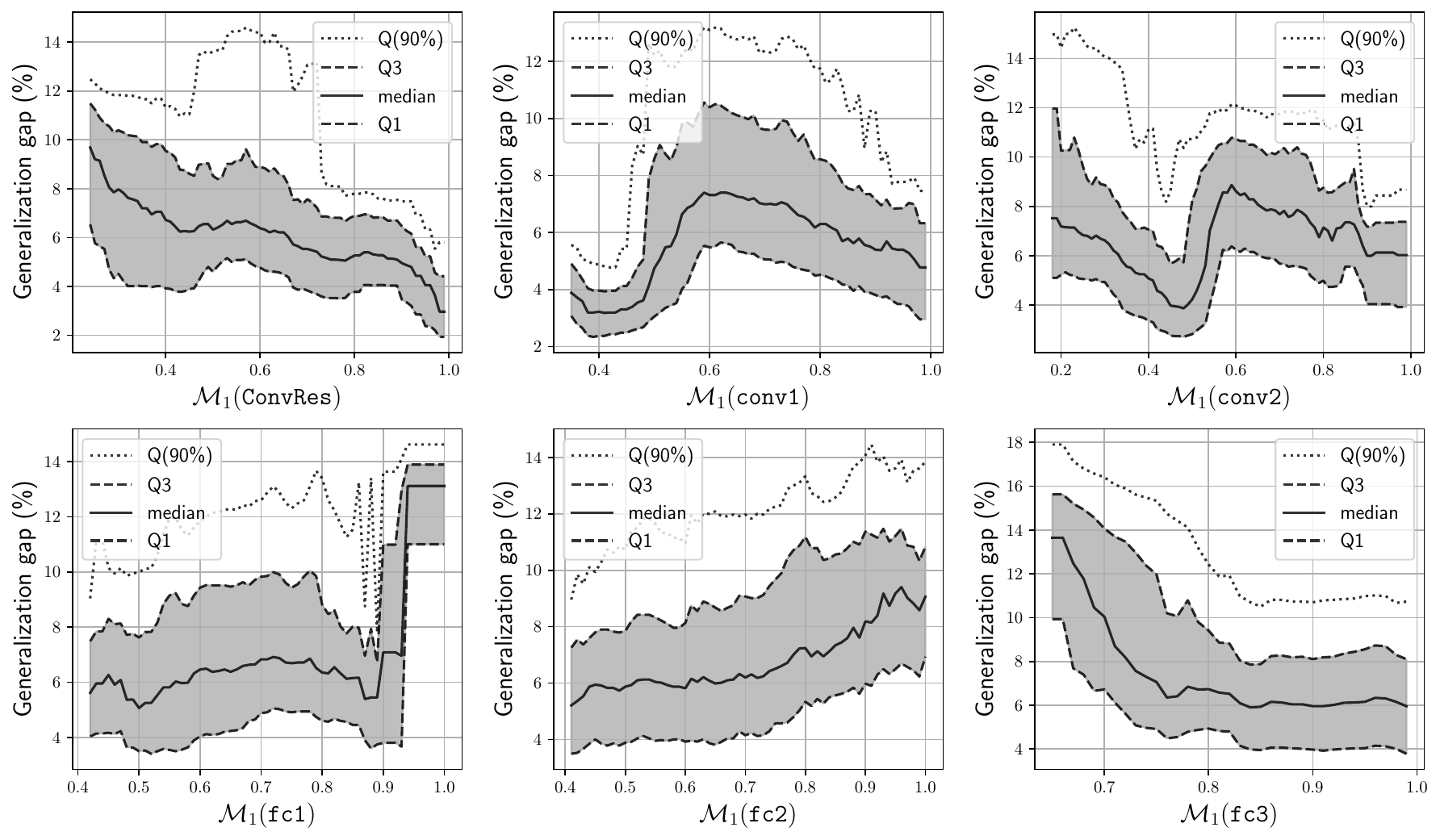}
    }
    \vspace*{-0.3cm}
  \caption{Quantile plots representing the evolution of the generalization gap over our 20 domains according to  $\mathcal M_1$ computed with latent margins relative to a single layer of our Bayar detectors. These metrics are normalized for comparison. Q1 is the first quartile, Q3 is the third quartile and Q(90\%) is the 90th percentile. Metric points are scanned with a step of 0.01 and a given window size of 0.1.}
  \label{fig:margin_gp_layer}
\end{figure*}
Until here, we used the margin distributions of each latent space of our Bayar detectors for the computation of $\mathcal{M}_{\alpha}$. 

Although \cite{margin} argues that examining the margins of a single latent space is inadequate for capturing the generalization gap, we still propose to investigate the correlation between $\mathcal G$ and $\mathcal M_{1}$ computing the margin statistics from each layer independently through Figure \ref{fig:margin_gp_layer}. This choice is guided by the observation of a small positive correlation between $G$ and $\mathcal M_1$ when we reach high margins that we would like to understand. We see that the expected correlation between $\mathcal G$ and $\mathcal M_{1}$ is clearly present using margin distributions from the very first and the very last latent spaces. However, this correlation does not hold for the intermediate latent spaces. We explain the correlation with the very first layer by the fact that upstreams layers are known to extract more general features than downstream layers in deep architectures \cite{upstream}. Hence, maximizing source margins at that level is also expected to maximize target margins at this same level, shrinking the final generalisation gap.
Concerning the very last layer, its the most specific layer of the architecture. Hence, if source margins are too tight at that level, noisy perturbation of the source (i.e. post-processings) lead to misclassifications. We believe this explains why a large class separation in this latent space is also beneficial for robustness against post-processing. This justify the recent trend towards the construction of a final latent space clearly separating the classes using contrastive losses ~\cite{contrastive}. Regarding the intermediate latent spaces, the absence of negative correlation between $\mathcal M_1$ and $\mathcal G$ is certainly because these layers are trained to separate classes in the final latent space without bothering to well separate them within their own latent space.

\subsection{Impact of parameters and operators on robust detection}

Here we propose to examine the impact of hyperparameters and operators on generalization gaps and the distributions of latent margins. For this analysis, we compute $\mathcal M_2$ using only the margins from first and last layers to enhance the correlation of our margin metric with the generalization gap.

We provide statistics helping to assess the impact of independently modifying hyperparameters and operators on the generalization ability of our detectors in Table \ref{tab:merged}. Our results demonstrate that $\mathtt{instance \ normalization}$ and $\mathtt{average \ pooling}$ are particularly effective choices for designing robust splicing detectors. We attribute this effectiveness through high median $\mathcal M_2$, showing that these operations significantly expand latent margins. Regarding hyperparameters, the most robust architectures are, as expected, those with the lowest median accuracy on the source. However, we must acknowledge that modifying $\mathtt{dropout}$ does not have a significant impact on robust detection, while modifying $\mathtt{batch \ size}$ has an impact that is not really explained by high margins. We believe this is because dropout and batch size are hyperparameters that have less impact on latent representations compared to normalization and pooling operators that respectively squeeze data and reduce its dimensionality.

% \begin{figure}[!h]
% 	\centerline{
% 	\includegraphics[width=\columnwidth]{imgs/network_withoutDA.pdf}
% 	}
% 	\caption{Architecture of our forgery detector largely inspired by \cite{bayar2016}, in red: \textcolor{red}{Batch Norm + ReLU}, in orange: \textcolor{orange}{Flatten}, in purple: \textcolor{Orchid}{ReLU + Dropout(0.5)}.}
% 	\label{fig:bayar}
% \end{figure}

% \section{Impact of non-learnable modules on Robust Detection}
% \label{sec:comparison}
% \input{4_comparison.tex}
\section{Conclusions and Perspectives}

\label{sec:conclusion}
This article explores how the robustness of a splicing detector against unknown post-processing can vary depending on its training. By examining factors that influence detector performance, we proposed several best practices for forensic analysts. Our research first showed that over-training a detector on a single source negatively affects its generalization to post-processed samples, prompting the need to determine optimal training stopping points. To help with this, we developed a margin metric correlated with the generalization gap by leveraging classical statistics to summarize latent margins distributions. Notably, we found that the first and last latent margins distributions significantly correlate with the detector's robustness against post-processing. Finally, we discovered that some pooling and normalization operators proved more effective than others in fostering post-processing robustness given the wide latent margins they produced. We currently recommend training splicing detectors with multiple hyperparameters choice and selecting the one maximizing margins in first and last layers. In future research, we plan to check the consistence of this correlation with different forgery detectors such as Noiseprint \cite{noiseprint} and Trufor \cite{trufor}. If the correlation persists, we propose to design architectures resilient to post-processing robustness by maximizing latent margins, leveraging for instance the contrastive losses that already proved effective in forensics \cite{contrastive}. We also intend to study the effects of different hyperparameter and operator combinations on out-of-distribution (OOD) generalization for splicing detectors.
%\newpage





\bibliographystyle{IEEEtran}


\bibliography{refs}

\end{document}